%% file: main.tex
\documentclass[conference]{IEEEtran}
\IEEEoverridecommandlockouts 
\usepackage{booktabs,amsmath,amssymb,amsfonts,mathtools,verbatim,xcolor,multirow,adjustbox,soul,breakcites,multirow}
\usepackage{float}
\usepackage{pdfpages}
\usepackage{url,graphicx}
\usepackage[linesnumbered,ruled,vlined]{algorithm2e} 
\usepackage[font=small,skip=7pt]{caption}
\usepackage{pbox}
\usepackage{cases}
\usepackage{graphicx}

\usepackage{amsmath}
\usepackage{mathtools}
\usepackage{makecell}
\usepackage{array}
\usepackage{bm}
\newcolumntype{C}[1]{>{\centering\arraybackslash}m{#1}}
\newcommand{\amt}{$amt$}

\newcommand{\intr}{$intr$}
\newcommand{\tp}{$tp$}

\newcommand{\MultiSig}{\mathsf{MultiSig}}
\newcommand{\OutReq}{\mathsf{OutReq}}

\begin{document}
\title{Balance Transfers and Bailouts in Credit Networks using Blockchains\textsuperscript{*}\thanks{\noindent\textsuperscript{*}Research supported by NSF award \#1800088. Any opinions, findings, and conclusions or recommendations expressed in this material are those of the author(s) and do not necessarily reflect the views of the National Science Foundation.}}

\author{
    \IEEEauthorblockN{Lalitha Muthu Subramanian, Roopa Vishwanathan, Kartick Kolachala}
    \IEEEauthorblockA{Department of Computer Science, New Mexico State University, USA
    \\\{lalitha,roopav,kart1712\}@nmsu.edu}
}
\maketitle
\input{abstract}
\input{introduction}
\input{relatedworks}

\input{sysdesign}

\input{advmodel}

\input{routing}
\input{algorithms}

\input{implementation}
\input{conc}


\bibliographystyle{IEEEtran}
\bibliography{references}
\end{document}

%% file: abstract.tex
\begin{abstract}
In this paper, we propose a technique for rebalancing link weights in decentralized credit networks. Credit networks are peer-to-peer trust-based networks that enable fast and inexpensive cross-currency transactions compared to traditional bank wire transfers, which has led to their increasing popularity and use. Although researchers have studied security of transactions and privacy of users of such networks, and have invested significant efforts into designing efficient routing algorithms for credit networks, comparatively little work has been done in the area of \emph{replenishing} credit links of users in the network. Replenishing links at regular intervals in a credit network is important to keep users solvent, the network viable with enough liquidity, and to prevent transaction failures. This is achieved by a process called \emph{rebalancing} that enables a poorly funded user to create incoming as well as outgoing credit links. 
\par
We propose a system where a user with zero or no link weights can create incoming links with existing, trusted users in the network, in a procedure we call \emph{balance transfer}, followed by creating outgoing links to existing or new users that would like to join the network, a process we call \emph{bailout}. Both these processes together constitute our proposed rebalancing mechanism. Our techniques would also serve to make the network more competitive by offering users lower rates of interest, and enable users to earn routing fees-based revenue by participating in high throughput transaction paths.

\end{abstract}

%% file: introduction.tex
\section{Introduction}
\label{sec:intro}
Blockchain and cryptocurrencies such as Bitcoin~\cite{nakamoto2008bitcoin} have disrupted the banking industry, enabled new business models, and helped in designing new, efficient financial infrastructure. A blockchain is an append-only distributed ledger, where users post messages or transactions, which are usually considered immutable. Blockchains have enabled the growth of IOU (I Owe You) credit networks in recent years. A credit network is a decentralized peer-to-peer lending network, where users lend out financial credit based on social trust. Credit networks provide the means to do path-based payments between two users, where the payment is routed through multiple intermediate users. Credit networks offer several advantages over traditional banks, such as low end-to-end transaction time, lower routing fees, and the ability to perform cross-currency transactions in the order of seconds. Some examples of real-world credit networks are Ripple~\cite{ripple} and Stellar~\cite{stellar}.
\par
Credit networks are usually modeled as a directed graph with vertices and weighted edges/links. The vertices represent the various users in the system and the weight on a link is the credit a user is willing to offer to an adjacent user. The directionality of the link is used to denote the lender and borrower, e.g.,  $u\overset{20}{\rightarrow} v$ denotes $v$ has extended 20 units of credit to $u$. Payments are routed along (and in the direction of) credit links, and once a payment is made from a sender to receiver, the link weights are decremented along the transaction path.
\par 
There has been growing interest in finding solutions for efficient routing, and enabling private and secure transactions in credit networks~\cite{prihodko2016flare,silentWhispers,panwar2019blanc} but not much work has been done in the area of \emph{rebalancing} link weights of a user that has run out of credit, and cannot participate in transactions. Rebalancing in credit networks is a significant problem to study, since, if the credit on a given link gets exhausted, i.e., if the weight on a link connecting two nodes drops to zero, no transactions can be done on any path containing those nodes until the link is refunded, a process which involves expensive on-chain transactions. Such nodes will be unable to participate in any transaction due to their inability to transfer or flow money, and will eventually be shunned by other nodes in the network. Networks which have large sections of such dormant nodes will eventually become inefficient, have progressively low throughput and may not even remain operational. In this paper, we study the problem of rebalancing links in an efficient way, while making minimal use of the blockchain, with the goal of avoiding mining and blockchain write fees.
\par
We propose a two-step rebalancing process wherein a node whose link weights are low or close to zero, can create fresh incoming links in a process called \emph{balance transfer}, and then create fresh outgoing links in a process called as \emph{bailout}. Upon completion of these two steps, the poorly connected node will become an active participant in the network, become involved in high throughput transactions, thus enabling it to collect routing fees. At a high level, balance transfer involves existing nodes moving outstanding debt to a new lender, incentivized by lower interest rates, and bailout involves the lender temporarily being assisted by a bank, who infuses capital into them to shore up their credit reserves. Once their earnings exceed their debts, the bank exits the system. 

\par
\textbf{Balance Transfer in Credit Networks}: Balance transfer in the real world occurs
when the outstanding balance of one credit card (or several credit cards) is
moved to another credit card account. This is often done by consumers looking
for lower interest rates. Many credit card issuers offer introductory balance
transfer APRs that are lower than the standard rates. Another advantage of balance transfer is that it makes financial management easier by transferring consolidated balances to
a new credit provider. Although lending in blockchain has become quite common, the idea of doing balance transfers in credit networks has not been explored. In this paper we
design a system where any lending node with low connectivity can advertise a lower interest rate, and thus gain more borrowers. 
\par
\textbf{Bailout in Credit Networks}: A bailout is a process where an organization or a government injects capital into a failing business to save it from bankruptcy, and to help make the business competitive again. In the context of credit networks, a bank does a ``bailout'' of a user, Alice, with no outgoing credit links by lending credit to her, and temporarily helps her by connecting her with other users with whom Alice can establish permanent links. Once she establishes permanent links with other users, the bank exits the network, possibly after collecting a small fee from Alice.
\par
\textit{Our Contributions}: In this paper, we give a new approach for rebalancing depleted credit links in credit networks.
\par
\noindent
1) We propose a two-step approach for rebalancing consisting of \emph{balance transfer} and \emph{bailout}. In the balance transfer step, a poorly connected node, called as \emph{requestor}, will establish incoming links with other nodes in the network by advertising a lower interest rate. This enables the requestor to become a lender to other nodes. In the bailout step, a well-known party such as a bank will infuse capital into the requestor node, by helping it connect to, and establish outgoing links with several other nodes in the network. After the requestor node establishes outgoing connections with other nodes, the bank will leave the network, possibly after collecting a fee from the requestor. At the end of this process, the requestor node will have several incoming and outgoing links, which will enable it to help facilitate several transactions, thus increasing the overall throughput and robustness of the credit network.
\par
\noindent
2) Since the performance of our balance transfer step is highly dependent on being able to find routes efficiently, we compare and analyze two different routing algorithms for doing balance transfers: Chord~\cite{chord} and VOUTE~\cite{roos2016voutevirtual}, and evaluate their performance experimentally.

%% file: relatedworks.tex
\section{Related Work}
\label{sec:related}
In this section, we  review literature on credit networks, payment channels, and rebalancing and loaning in credit networks.
\par
\noindent
\textbf{Credit networks}: Fulgor and Rayo~\cite{concurrency} were the first to setup a peer-to-peer payment channel network that provides provable security and privacy properties, with Rayo being the first payment network that enforces non-blocking transactions. Fulgor and Rayo, both, establish a path between sender and receiver, assuming all users in the path to be honest, and users have at least partial knowledge about network topology. Unlike Fulgor and Rayo, in our system, the entire network topology is not known to the users and we do not assume all the users in the path to be honest. Also, we do not propose any payment operations, instead focus only on rebalancing credit links. 
\par
SilentWhispers~\cite{silentWhispers} presents a decentralized credit network (DCN) architecture which consists of subsets of paths between the sender and receiver calculated via several trusted entities called \emph{landmarks}. PathShuffle ~\cite{pathshuffle} presents a path mixing protocol for Ripple network providing complete  anonymity. PathShuffle leverages existing infrastructure from Ripple~\cite{ripple} to create and maintain consistent credit links with users in the networks. Both~\cite{silentWhispers} and~\cite{pathshuffle} present solutions for routing payment in a secure and privacy-preserving way in credit networks, and neither tackle the challenge of rebalancing depleted credit links. Our goal is to design a mechanism for rebalancing credit links.
\par
We leverage the concept of \emph{landmark} nodes proposed in~\cite{silentWhispers} in our system to assist the requestor node in our bailout phase. The landmark node is a well connected node such as a bank, and hence, can potentially help the requestor node contact several other nodes in the network for establishing outgoing links. We use the landmarks in our system to assist the requestor node establish outgoing links without placing enormous trust on landmarks, unlike~\cite{silentWhispers}. Roos et al~\cite{PBTsettling} used graph embedding for efficient routing with concurrent transactions overcoming some inefficiencies in~\cite{silentWhispers}. They too do not focus on the rebalancing problem. Malavolta et al~\cite{dlsag}, recently proposed a novel linkable ring signature scheme for refund transactions natively in Monero~\cite{monero} and extend the same scheme into having scalable off-chain transactions by establishing payment channels using Monero. Our system could be deployed in such kind of payment channel network to enable the system to be more competitive and enhance the connectivity in such channels simultaneously achieving rebalancing within the network. 
\par
Panwar et al.~\cite{panwar2019blanc} proposed a DCN system where users can perform path-based transactions that preserves sender, receiver and value privacy but involves a significant number of blockchain writes in the course of a normal, successful transaction (more in the case of transaction rollbacks, re-tries, and other edge cases). Our system also involves blockchain writes, but in our system, a single blockchain write is done only after the completion and execution of the entire rebalancing protocol, unlike~\cite{panwar2019blanc}, which would help it scale much better.
\par
\noindent
\textbf{Rebalancing}:
REVIVE~\cite{khalil2017revive} is a payment network that allows users to rebalance their channel without having to communicate with the blockchain.  Although very efficient in rebalancing bidirectional networks (i.e., cyclic networks), REVIVE does not present any solution for rebalancing in a unidirectional (acyclic) credit network, which is our use-case scenario. REVIVE has a leader elected in the network who stays online all the time in order to facilitate rebalancing requests. We do not place trust in any leader or centralized entity to establish new incoming or outgoing links in the system although we make use of landmark nodes to assist the requestor node \emph{temporarily} in our bailout step. Lightning network~\cite{poonlight} is a highly scalable payment channel network that is constructed on top of Bitcoin. Lightning network does re-balancing off-chain, but again only for \emph{cyclic} networks. To rebalance a credit link, a node does a payment in cyclic path to itself and such payments usually comes with a fee for each node in the circular path. In our approach, a node establishes new incoming and outgoing credit links, without having to pay every node in a payment path.
\par
Ripio Credit Network (RCN)~\cite{Ripio} is a peer-to-peer global credit network protocol based on cosigned smart contracts and blockchain technology. A user can join as a lender, cosigner or borrower in the network. A cosigner will act as a go-between a borrower and a lender in the network. In case a borrower defaults, the cosigner works out an alternative mechanism for managing debt collection. However, this places a great responsibility on the cosigner, and if they are incapable of actually enforcing debt collection for any reason, the overall value of the network would decrease.



%% file: sysdesign.tex
\section{System Design}
\label{sec:system}

Credit networks are usually dense networks, e.g., Ripple~\cite{ripple}, with several incoming and outgoing links from the nodes. If a node has depleted credit links, then, intuitively, one way for it to rebalance its links would be to extend credit to, and borrow from new users. This could be problematic for several reasons: the new users might not be trusted, or at the moment when a node's credit links are depleted there might not be enough new users in the system. Ultimately, whether to lend or borrow from a user, we believe, should be a matter of choice, and no new node should be compelled to accept credit from, or lend credit to an existing user, simply because the existing user needs to rebalance their credit links. With this design goal in mind, we introduce the concept of balance transfer and bailout where any existing node can transfer its credit links to a new lender who offers a lower rate of interest, and an existing use can actively look for credit lenders in the system, with some help from a partially-online trusted bank. We now give an overview of the two steps that comprise our system.
\par
\begin{figure} 
\centering
 \includegraphics[scale=0.4]{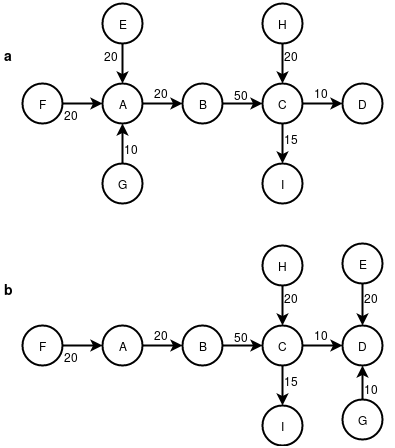}
\caption{Balance transfer in credit networks}
\label{fig:baltrans}
\end{figure}

\noindent
\textbf{Balance transfer}: Figure~\ref{fig:baltrans} shows how the balance transfer process takes place in a credit network. Here any user can disconnect from an existing lender and transfer credit links to a new lender node offering a lower rate of interest.
 
\par 
In Figure~\ref{fig:baltrans} Part a) denotes a simple credit network where $A$ is a highly connected node. $D$ is a poorly connected node and would need to send requests to establish new incoming connections. To this end, $D$ first raises a request to add incoming nodes, by advertising a lower rate of interest, which will incentivize some of the nodes in the network to transfer their credits from their existing lender to $D$. It is important to note here that, in our design, every node that wants to transfer to $D$ would have to transfer their full credit that exists with the current lender, and nodes are not allowed to establish an incoming link to $D$ with partial amount. This closely models the real-world balance transfer mechanisms where a user either transfers their debts (to a new lender) in the debt's entirety, or not at all, but cannot partially transfer debts. Figure~\ref{fig:baltrans} Part b) depicts the network after two of $A$'s borrowers, $E, G$ have voluntarily transferred their balances to node $D$, after severing their links with $A$, thus $D$ has established several incoming links. At this point, all affected nodes will locally store their new links and link weights. After the bailout step, the changed network topology will be written to the blockchain.
\par
\noindent
\textbf{Bailout}: In the bailout process, a trusted, highly connected party such as a bank, or a credit union \emph{temporarily} lends credit to node $D$, so that $D$ can establish outgoing connections. We refer to this trusted party as a \emph{landmark}, or $LM$. The high-level idea is that $LM$ will use the fact that is is highly connected, and temporarily connect $D$ with several other nodes in the network with whom $LM$ has a direct connection. $D$ will then request each of these nodes if they would like to lend credit to $D$, thus establishing outgoing links form $D$ to them. Note that any or all nodes can decline $D$'s request, at which point $LM$ will connect $D$ with a fresh set of nodes. 
\par
\begin{figure} 
\centering
 \includegraphics[scale=0.4]{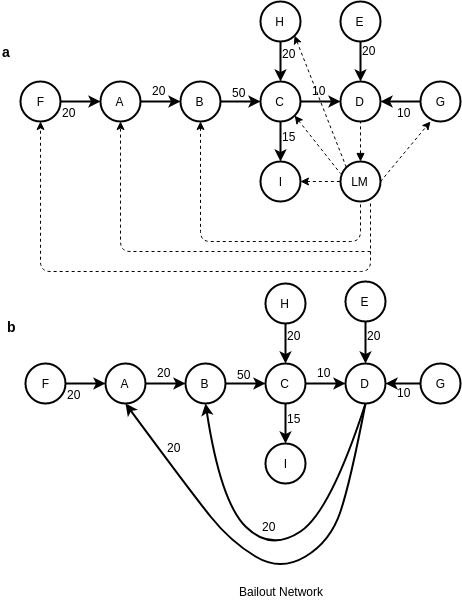}
\caption{Figure showing before and after scenario of the bailout step.}
\label{fig:bailout}
\end{figure}

Figure~\ref{fig:bailout} Part a) shows the bailout phase where the link between $LM$ and other nodes are established after the balance transfer state. At this point, the poorly connected node $D$ has two incoming links after the execution of the balance transfer step. In order to make $D$ an active participant in transactions in the credit network, $D$ also needs outgoing links. After the balance transfer scenario, node $D$ sends request to the highly connected node, $LM$. $LM$ provides a list of nodes, $F, A, B, I, C, H, G$, that can potentially establish a outgoing link with $D$, i.e., willing to lend credit to $D$. $D$ will then send request to all of these nodes for an outgoing link. If none of the neighboring nodes are willing, $LM$ helps $D$ to establish an outgoing link with any of the newer nodes that joined the system, or will give $D$ a fresh list of nodes. Figure~\ref{fig:bailout} Part b) shows the network after nodes $A$ and $B$ agree to $D$'s request for outgoing links, $D$ establishes outgoing links with $A, B$, at which point $LM$ exits the network.

%% file: advmodel.tex
\section{Adversarial Model}
\label{sec:adv}

In our system, we assume that any adversary can corrupt a single or a set of users in the network. The corrupted user(s) can be either the requester, who raises a rebalancing request, the nodes that respond to the request, or any intermediate node. During the bailout phase, we need a trusted landmark, $LM$, who is temporarily assists the requestor node. We assume the adversary cannot corrupt the $LM$ node. Each user $i$ has her own signing and verification key pair $(sk_{i},vk_{i})$ and an encryption, decryption key pair $(pk_{i}, dk_{i})$. Once any user is corrupted, their corresponding signing and verification keys are compromised, the adversary can misreport the credit link between a user and her neighbor, not respond to any request, respond selectively to requests and relay fraudulent balance transfer requests to its neighbors. We assume that an adversary cannot corrupt all users in the network, and thus may know partial network topology, but does not know the entire network. We now give the desired security/privacy properties of a payment network that enables balance transfers and bailouts. 

\subsection{Privacy and Security properties}
\noindent
\textbf{Link privacy}: Link privacy is achieved when an adversary only knows the value of links adjacent to her and will not have access to other nodes' links, even if they were part of the balance transfer or bailout steps. 
\par
\noindent
\textbf{Corrupted users}: We now discuss which users could possibly get corrupted, what can a corrupted user do, and how to mitigate the situation.
\par
\noindent
1) \textit{Corrupt balance transfer requestor node}: Any user who sends the balance transfer request and can act maliciously, by claiming to have committed to one value (credit limit) in the request tuple, and later reneging on their commitment. The requestor node could also refuse to provide a low rate of interest, as advertised. 
\par
\noindent
2) \textit{Corrupt responder node}: Any node that responds to the balance transfer request can act maliciously by claiming to have responded with a different amount than the actual amount linked to the request id. 
\par
\noindent
3) \textit{Corrupt intermediaries}: Any user along the route from requestor to responder can be a malicious. Such adversarial users can either drop messages with or without making an entry into their logs, or mis-direct messages to other collaborating nodes in the network. 
\par
\noindent
\textbf{Accountability}: Any malicious user should not be able to misreport her link value. In our system, each user maintains a record of their link weights with their next-hop neighbors in a local hash table. Each user involved in a rebalance transaction also holds signed contracts containing the current and updated link weights as a proof of link weight update. In case where any user behaves maliciously, the honest peers would be able to detect such malicious activity, also third-party arbiters can adjudicate based on the signed contracts. An arbitrator can be any law enforcing authority who on receiving a complaint can enforce legal punitive action according to local laws, or revoke nodes' access to network. The exact nature of the remedial action taken by arbitrators is beyond the scope of this paper.

%% file: routing.tex
\section{Routing in Balance Transfers}
\label{sec:routing}
\par \noindent
In this section we discuss two different routing protocols we use for balance transfers in credit networks: prefix embedding~\cite{greedy} and Chord~\cite{chord}, and analyze and compare their efficiency. Note that these routing algorithms are not used for routing payments among nodes in our system, rather only for doing a balance transfer. 

\subsection{Routing using Prefix Embedding}
The first part of the balance transfer algorithm is the \textit{Find Route} phase in which the responder node finds a route from itself to the requester. 
We make use of prefix embedding~\cite{greedy} and VOUTE~\cite{roos2016voutevirtual} to establish a route after which rebalancing occurs. Prefix based embeddings are a part of greedy embeddings~\cite{greedy}~\cite{greedyR}. They are, in general, created by embedding a spanning tree into a suitable metric space. An ID is assigned to the root node, and the tree consists of several child nodes where each child computes the ID based on the ID of its parent node. Prefix embedding is an adaptation of PIE embedding for unweighted graphs. The idea is that, every node is assigned an ID using a custom metric space such that the node ID is equivalent to the hop distance or the depth of a spanning tree. A child's id is essentially the ID of the parent, an additional coordinate equal to the index of the child. The prefix embedding algorithm uses the following equation to find the length of the shortest path between node $u$ and node $v$, where $cpl$ is the common prefix length. 
\begin{equation}
d(id(u),id(v)) = {id(u)} + {id(v)} -2cpl(id(u),id(v))
\label{eqn:eqn1}
\end{equation}

\subsection{Routing using Chord}
Chord~\cite{chord} is a routing algorithm built using distributed hash tables for peer-to-peer networks, without any centralized monitoring authority. Chord uses a $\langle key, value \rangle$ pair to map to a specific node across the system. The keys are assigned to nodes using Consistent Hashing~\cite{CH} across the network. Consistent hashing in Chord reduces the load in the system, since each node in a network requires the same number of keys and requires little movement of keys when nodes leave or join the system, making the system dynamically compatible.
\par \noindent
Chord assigns a $\langle key, value \rangle$ pair to each of the nodes in the system, where the key is the identifier using SHA-1~\cite{hashs} and maps the keys to the nodes that are responsible for them. A peer identifier is chosen by hashing the data key. The length of the identifier is usually large to ensure the probability of keys hashing to the same identifier is negligible. The identifiers are arranged from $0$ to $2^{m}-1$, where $m$ is the digest size of the hash function used. Key $k$ is assigned to the first peer whose identifier is equal to or follows $k$ in identifier place and the first peer, clockwise from $k$ is called the successor peer of $k$, represented by $successor(k)$. When a peer $n$ joins or leaves the system, the keys that previously belonged to $n$, is reassigned to $n's$ successor. This enables maintaining consistent hashing in the system. For helping users join and leave the Chord network, we run a \emph{stabilization} protocol at regular intervals that updates the \emph{finger table} stored at each node. The finger table (FT) stored at each node is a table containing the IDs of its successors. Due to space constraints, we do not elaborate further on the working of prefix embedding and Chord; below we briefly compare their efficiency in terms of routing efficiency and network restrictions. 

\subsection{Prefix Embedding vs. Chord}
\textit{Routing efficiency}: The main advantage using a Chord-based routing algorithm is that the number of hops to receiver is reduced, based on the density of the network. A peer leaving or joining the system does not involve too many changes to the key distribution to the system, although the successor pointers of some peers need to be changed. In prefix embedding, although, the users finds the shortest path, the worst case scenario for routing efficiency would be to find a route that traverses along the entire depth of the network.
\newline
\textit{Network Restrictions}: A peer leaving or joining the system does not involve too many changes, although the successor pointers of some peers need to be changed. It is important to ensure that the successor pointers are up to date otherwise, the routing will fail in such a system. Hence there is a need to constantly update the finger table, as and when a change occurs. When peers fail, it is possible that a peer does not know its new successor, and that it has no chance to learn about it. Hence, the efficiency at which the finger tables are updated are $O(\log^{2} N)$ ($N$ is the number of nodes). In prefix embedding or VOUTE, the network need not monitor their nodes constantly and there is no maintenance cost incurred. The users can join and leave when they want making the network more adaptable and does not involve handing over keys, updating peers about leaving the system etc.

%% file: algorithms.tex
\section{Construction}
\label{sec:cons}

In this section, we describe the construction of our system, comprising of the balance transfer and bailout steps. We first present two different ways of doing the balance transfer step: using prefix embedding-based routing, and using Chord-based routing. Then we present our algorithms for the bailout step.


\subsection{Balance Transfer using Prefix Embedding}
In this step, a responder node responds to a balance transfer request $BT$ broadcast by another node in the network. The requester node broadcasts the amount available for accommodating the incoming nodes and the rate of interest. Any user $j$ who wants to transfer to user $i$, needs to find a route to $i$, and then initiate the balance transfer. This process is depicted in Algorithm~\ref{alg:prefixembedding}.
\par
In Algorithm~\ref{alg:prefixembedding}, the parties involved are the requestor node, $i$, the responder node, $j$, and other intermediate nodes along the length of the response path. Node $i$ raises a balance transfer request by broadcasting the tuple $BT$. The tuple consists of the amount that $i$ can offer as credit, interest rate $intr$ and a response time $tp$ until when $i$ can accept responses from different nodes in the system (line 1,2). In line 3, $j$ finds the length of the route path $\mid L \mid$ using prefix embedding, where $\mid L \mid= \mid \overrightarrow{id_j} \mid + \mid \overrightarrow{id_i} \mid - 2 *cpl (\overrightarrow{id_j},\overrightarrow{id_i})$. Node $j$ finds the nearest node on the route, and computes hash of it's co-ordinates $\overrightarrow{id_j}$ in line 5. This hashed co-ordinates are used in prefix embedding to find the next neighbor and is computed using the common prefixes between $j$ and next node $k$. Node $j$ computes $\overrightarrow{id_k}$  in line 7, and  creates a signature  $Sign_{sk_{j}}(id_k) \longrightarrow \sigma_{j}$. Node $j$ does a $j.write(\overrightarrow{id_k})$ and $j.write(\sigma_{j})$ in line 8 of this algorithm. By using hashed coordinates, the address of the users are not made public, and the actual addresses are only available to the next-hop neighbors on the path. Then, $j$ sends a response $resp_{j} = \langle E_{pk_i}\left(amt_j\right)\rangle$ to $k$. In line 11-13, $j$ waits to ensure that $k$ does a $k.write$ of next address along the path, which ensures that the response is sent to the hashed address that is written to the shared hash-table between $k$ and its neighbors. If $j$ finds a malicious entry on the shared hash-table, $j$ can choose to send the response $resp_j$ through a different route. 
Node $i$ receives responses $resp_j$  in line 15 of this algorithm. Node $i$ saves all responses in $i.resp[]$ if the responses are received within time $tp$ and the amount in response $amt_j \leq amt$ in lines 16, 17. In line 19 of this algorithm, source node $src$ is set to $k$, $user = j$, and $dest = i$. For every response that $i$ receives, $i$ verifies if updated $amt \neq 0$, and calls $\MultiSig$ algorithm in line 21 and produces contract $CtBal$. The $\MultiSig$ function creates shared, signed contracts between two adjacent nodes; we do not give the algorithm here due to space constraints. Once $i$ initiates the $\MultiSig$ algorithm, node $j$ pays the amount lent by $k$ in line 22. Then, node $j$ updates her hash-table after receiving $k.resp$ and deletes the hash-table entry with $k$. Node $k$ deletes her entry in  hash-table and they sign an acknowledgement of the update (line 24, 25), and produces $Sign_{sk_{src}}(ack) \longrightarrow \sigma_{ack_{src}}$. The $\MultiSig$ operation returns $\lbrace{CtBal, \sigma_{dest},\sigma_{user}\rbrace}$, which is written to blockchain by $i$ in line 26 of this algorithm.

\begin{algorithm}[htbp]
\DontPrintSemicolon
\SetAlgoLined
\SetKwInOut{InitialState}{Initial State}
\SetKwInOut{FinalOutcome}{Final Outcome}
\SetKwInOut{Parties}{Parties}
\InitialState{Node $i$ requesting for Balance transfer}
\FinalOutcome{Nodes $j$ establish a route to $i$ for balance transfer. }
\Begin
{
\tcc {Find Route Phase}
$i$ raises Balance Transfer request by broadcasting tuple $BT$ = $\langle \amt,\intr,\tp \rangle$\\
$j$ the receiver computes $\mid L \mid= \mid \overrightarrow{id_j} \mid + \mid \overrightarrow{id_i} \mid - 2 *cpl (\overrightarrow{id_j},\overrightarrow{id_i})$\\
\For {$j \in \mid L \mid$}
{
$j$ computes hashed co-ordinates $\overrightarrow{id_j}$ with padding added\\
$j$ finds the id of next node $k$, using common prefix coordinates\\
$j$ hashes $\overrightarrow{id_k}$ and creates signature $Sign_{sk_{j}}(id_k) \longrightarrow \sigma_{j}$\\
$j$ does $j.write(id_k)$ and $ j.write(\sigma_{j})$\\
$j$ creates response $resp_{j} = \langle E_{pk_i}\left(amt_j\right)\rangle$ and sends  to $k$\\
$j$ waits for time $t_j$, verifies if node $k$ does $k.write$\\
\If {$k.write = ``False"$}{
$j$ finds a different route until $resp_j$ reaches $i$\\
}
}
\tcc {Balance Transfer Phase}
Node $i$ receives $resp_{j}$ \\
\If {$ts <tp$ and $amt_j \leq amt$}
{
Node $i$ accepts  and saves responses in $i.resp []$ \\ 
}
Set $src =k$ , $user =j$, $dest =i $\\
\For {each $user \in i.resp$}{
$dest$ node calls $\MultiSig(dest = i,val,user)\longrightarrow (CtBal, \sigma_{dest},\sigma_{user})$\\
$user$ pays back $lw_{src,user}$ and creates $req_{(user,src)} $ with $lw_{src,user} = 0$\\
$src$ signs $ack$ of $lw_{src,user} = 0$, produces $Sign_{sk_{src}}(ack) \longrightarrow \sigma_{ack_{src}}$\\
$user$ verifies  $\sigma_{ack_{src}}$ and creates $\sigma_{ack_{user}}$\\
$src$ and $user$ deletes the hashtable entries on $src$ and $user$ nodes respectively\\
$dest$ node writes $\lbrace{CtBal, \sigma_{dest},\sigma_{user}\rbrace}$ to blockchain\\
}
}
\caption{Balance Transfer Algorithm using Prefix Embedding}
\label{alg:prefixembedding}
\end{algorithm}



\subsection{Balance Transfer Algorithm using Chord}
In this section, we discuss the balance transfer process using Chord for routing, Algorithm~\ref{alg:FRchord}. The first phase of this algorithm consists of the Find Route phase, where the user $j$, referred to as the responder node,  responds to a $BT$ request raised by a requestor node $i$. Node $i$ raises a request in line 2 of this algorithm with a tuple $BT$ = $\langle amt,intr,tp,key \rangle$, where, $amt$ is the amount that is extended as credit, $intr$ is the rate of interest offered, $tp$ is the time period within which the response is accepted and $key$ which is used in finding the route to the user in a chord network. Node $i$ broadcasts the request $BT$ to all users in line 3. Node $j$ responds to $BT$ by first locating the $key$ of $i$, in line 4. If node $i$ is an immediate neighbor of $j$ from its finger table $FT_{j}$, $j$ responds with $resp_{j} = \langle E_{pk_i}\left(amt_j\right)\rangle$ in line 6. If node $i$ is not among $j$'s immediate neighbors, $j$ finds the nearest node $k$ from $FT_{j}$ and sends the response  in line 9 of this algorithm. Every intermediate node $k$ along the length of the path $L$, does a lookup(key) in their corresponding $FT_k$ in line 11.  Every node $k$ does the steps 6 to 13 until the $resp_j$ reaches node $i$ (requester). Node $i$ receives all responses accepted within time $tp$ and stores in $i.resp[]$ in line 16. In line 18, source node $src$ is set to $k$, $user = j$, and $dest = i$. For every response that $i$ saved, $i$ verifies if updated $amt \neq 0$, calls the Multisig algorithm in line 20, produces contract $CtBal$. Each $user$ node in $i.resp[]$ then updates the link weight between $user$ and $src$ to 0 in line 21. The $src$ node signs $ack$ of $lw_{src,user} = 0$, produces $Sign_{sk_{src}}(ack) \longrightarrow \sigma_{ack_{src}}$ in line 22 of this algorithm, $user$ verifies $\sigma_{ack_{src}}$ and creates $\sigma_{ack_{user}}$ in line 23, and in line 24, the nodes $src$ and $user$ delete their corresponding hash table entry. The $dest$ node writes $\lbrace{CtBal, \sigma_{dest},\sigma_{user}\rbrace}$ to blockchain in line 25. Once the MultiSig operation is complete, node $j$ repositions itself in the Chord ring after informing the successors and predecessors in the Chord $FT$ in line 27 and 28. Every node from $j$'s previous position to new position, update their finger table by calling the $stabilize(FT)$ function to update their respective finger table entries. $j$ also calls the $Stabilize$ function to update her finger table with new successors after relocating in line 32. 

\begin{algorithm}[htbp]
\DontPrintSemicolon
\SetAlgoLined
\SetKwInOut{InitialState}{Initial State}
\SetKwInOut{FinalOutcome}{Final Outcome}
\SetKwInOut{Parties}{Parties}
\InitialState{Node $i$ requesting for Balance transfer}
\FinalOutcome{Nodes $j$ establish a route to $i$ for balance transfer. }
\Begin
{
\tcc {Find Route Phase}
Node $i$ raises Balance Transfer request with tuple $BT$ = $\langle \amt,\intr,\tp,key \rangle$, where $i \in N$\\
$i$ broadcasts $BT$ to all users\\
Any $j \in N-i$ does a $lookup(key)$ in her finger table $FT_j$\\
\If{$i \in FT_j$}{
Node $j$ responds to $BT$ with $resp_{j} = \langle E_{pk_i}\left(amt_j\right)\rangle$ and sends to $i$\\
}
\Else{
Node $j$ finds nearest node $k$ from $FT_j$ and sends response $resp_{j} = \langle E_{pk_i}\left(amt_j\right)\rangle$ to $k$\\
\For{each $k \in L$, where $L$ is the length of the path}
{
$k$ does a $lookup(key)$ in her finger table $FT_k$\\
Node $k$ does steps 6 to 13 until $resp_{j}$ reaches $i$\\
}
\tcc {Balance Transfer Phase}
Node $i$ receives $resp_{j}$ \\
\If {$ts <tp$ and $amt_j \leq amt$}
{
Node $i$ accepts  and saves responses in $i.resp []$ \\ 
}
Set $src =k$ , $user =j$, $dest =i $\\
\For {each $user \in i.resp$}{
$dest$ node calls $\MultiSig(dest = i,val,user)\longrightarrow (CtBal, \sigma_{dest},\sigma_{user})$\\
$user$ pays back $lw_{src,user}$ and creates $req_{(user,src)} $ with $lw_{src,user} = 0$\\
$src$ signs $ack$ of $lw_{src,user} = 0$, produces $Sign_{sk_{src}}(ack) \longrightarrow \sigma_{ack_{src}}$\\
$user$ verifies  $\sigma_{ack_{src}}$ and creates $\sigma_{ack_{user}}$\\
$src$ and $user$ deletes the hashtable entries on $src$ and $user$ nodes respectively\\
$dest$ node writes $\lbrace{CtBal, \sigma_{dest},\sigma_{user}\rbrace}$ to blockchain\\
} 
Node $j$ informs $successors, predecessors$ about re-positioning\\
Successors of node $j$ update their corresponding $FT$\\
Successors call $Stabilize(FT)$ to update their finger tables\\
$j$ hands over $key_j$ to $j$'s successors\\
$j$ re-positions in Chord table, re-assigns nodeID\\
$j$ calls $stabilize(FT_j)$ \\
$j$ updates $FT_j$ \\

}
}
\caption{Balance Transfer Algorithm using Chord}
\label{alg:FRchord}
\end{algorithm}

\subsection{Bailout Phase}
The bailout step is the final step in the rebalancing process and is given in Algorithm~\ref{alg:bailout}. Node $i$, in this algorithm requests for outgoing nodes through the $LM$ node. The main aim of this algorithm is to connect $i$ with multiple outgoing nodes successfully with the help of the landmark node ($LM$). In line 2 of this algorithm, node $i$ sends a request to $LM$ to connect or find $m$ nodes that would ideally, like to establish an outgoing link with $i$. $LM$ returns a list of $j$ identities, where $j \in [1,2,..m]$. $LM$ waits for a response time $tr$ within which atleast one outgoing link is established and this is done in line 4. In line 6, $i$ calls the $OutReq \longrightarrow val$ to $j$ through $LM$. In line 7 of this algorithm, if any $j$ responds with a ``val", then $j$ calls the $\MultiSig(dest = j, user =i, val) \longrightarrow (CtBal, \sigma_{dest},\sigma_{user})$ function. If all $j$ nodes respond with a ``$\perp"$, then $LM$ responds with a new set of $j \in [v_1,v_2,v_3...v_m]$ in line 11 of this algorithm. In line 13, $j$ calls $\MultiSig(dest = j, user =i, val) \longrightarrow (CtBal, \sigma_{dest},\sigma_{user})$. In line 16, $i$ writes $(CtBal, \sigma_{dest},\sigma_{user})$ to blockchain. In line 17, $LM$ exits the network after collecting a small fee from $i$, which is computed based on the number of links established. In the unlikely event that none of the nodes in the list sent by $LM$ is interested in establishing a outgoing connection from $i$, $LM$ does steps 2-13 again with a new list of $j$.
\begin{algorithm}[htbp]
\DontPrintSemicolon
\SetAlgoLined
\SetKwInOut{InitialState}{Initial State}
\SetKwInOut{FinalOutcome}{Final Outcome}
\SetKwInOut{Parties}{Parties}
\InitialState{Node $i$ requesting for outgoing credit links}
\FinalOutcome{Node $i$ establishes outgoing links with $j$ nodes }
\Begin
{
$i$ sends $m$ outgoing nodes request to $LM$.\\
$LM$ creates links $LM \longrightarrow j$, $j \in [1,2...m]$\\
$LM$ waits for $tr$ for $i$ to respond with outgoing request\\
\While{$tr \neq 0$}{
$i$ calls $OutReq(i,j,Inlink,t) \longrightarrow val$ to $j$ through $LM$, $t<tr$\\
\If {$j.resp$ =``val"}{
$j$ calls $\MultiSig(dest = j, user =i, val) \longrightarrow (CtBal, \sigma_{dest},\sigma_{user})$\\
}
\If{ $j.resp = ``\perp"$, for all $j \in [1,2...m]$}{
$LM$ responds with new nodes $\langle v_1,v_2,v_3..,v_m \rangle$\\
$i$ calls $OutReq(i,j,Inlink,t) \longrightarrow val$ to $j$,$j \in [v_1,v_2...v_m]$ through $LM$, $t<tr$\\
$j$ calls $\MultiSig(dest = j, user =i, val) \longrightarrow (CtBal, \sigma_{dest},\sigma_{user})$\\}
}
$i$ writes $(CtBal, \sigma_{dest},\sigma_{user})$ to blockchain.\\
$LM$ disconnects from $i$ after collecting her fee \\
}
\caption{Bailout Algorithm}
\label{alg:bailout}
\end{algorithm}

%% file: implementation.tex
\section{Implementation and Evaluation}
\label{sec:impl}
We simulated Chord-based routing using~\cite{peersim} as a base, and implemented cryptographic primitives in Charm~\cite{charm}. For signatures, we used ECDSA on curve prime192v2 and used RSA 2048 for encryption. Our experiments were run on a desktop class computer with Intel(R) Core(TM) i3-7100 CPU @ 3.90GHzx4 and 8GB RAM on Ubuntu-18.04 platform. Table ~\ref{tbl:multisigtime} shows time taken for cryptographic operations in every phase. 
\begin{table}[htbp]
\centering
\caption {Time for cryptographic Operation by Phase} 
\label{tbl:multisigtime}
\begin{tabular}{p{2cm}|p{1.5cm}|p{1.5cm}|p{1.5cm}}
\hline
\hline
 Operation & Signature (msec) & Verification (msec) & Encryption (msec) \\
\hline\hline
Balance Transfer (Prefix Embedding) &3.355 &3.7733 & 0.1707\\
\hline
Balance Transfer (Chord) &1.9678 & 2.2786 & -\\
\hline
Bailout &3.355 &3.7733 & - \\
\hline\hline
\end{tabular}
\end{table}
\par
In order to simulate the chord network, we have made use of \cite{peersim} simulator. The experiment was run on an intel i-3 generation-7 desktop with 8GB of RAM.  We recorded the time taken for setting up the chord ring, broadcasting a message and we also calculated the minimum and maximum number of hops it takes for a source to reach its destination.  In the ``Setup'' operation, we record time taken for setting up the Chord network ring for a total of 1000 nodes. The ``Lookup'' operation involves finding the route by performing a $lookup(key)$ from the finger table. Time taken for response phase is the total time taken for doing a $lookup(key)$ and respond to the broadcast message. ``Re-assign and Stabilize'' is the final operation which involves updating finger tables for all the successors and predecessors of the re-positioned node, after the balance transfer operation has been carried out. The time taken for setup varies linearly with the number of users added to the Chord network.  

\begin{table}[htbp]
\centering
\caption {Balance Transfer in Chord - Time recorded for operations} 
\label{tbl:chordtime}
\begin{tabular}{p{4.5 cm}|p{2cm}}
\hline
\hline
 Operation & Time taken \\
\hline\hline
Setup (1000 users) & 23.0650 sec\\
\hline
Broadcast BT & 13.913 sec\\
\hline
Lookup and Response  & 20.7730 sec \\
\hline
Re-assign and Stabilize & 10.6676 sec \\
\hline\hline
\end{tabular}
\end{table}
\par
We give the timings for balance transfer using prefix embedding in Table ~\ref{tbl:PIEtime}. We used the GTNA package~\cite{gtna} to implement balance transfer using Prefix Embedding. The time taken for ``Setup'' which involves, creating a network of 1000 users, assigning links and link weights and creating  embedding co-ordinates is given in table ~\ref{tbl:PIEtime}. The ``Broadcast BT''  represents the time taken for a requester sending out a BT (Balance Transfer) request tuple, which contains the position and amount for balance transfer. ``FindRoute and Respond'' operation is the process of finding route between a requester and responder, and also includes the time taken for sending out the response which is the encrypted value of the amount that any user is willing to balance transfer with the requester. This also includes the time taken for the user to establish an edge with the  requester (create an edge between two nodes in the graph). The setup time increases linearly with increase in number of users, as expected. For ``Find Route and Response'', the time taken to route from the first node to the farthest node in the graph is given here, since that would be the longest routing path, a node would encounter. Since the graph generated by GTNA is unstructured, and randomly generated each time, we give the average value of 100 runs for 1000 users in Table~\ref{tbl:PIEtime}. 

\begin{table}[htbp]
\centering
\caption {Balance Transfer using Prefix embedding - Time recorded for operations} 
\label{tbl:PIEtime}
\begin{tabular}{p{4.5 cm}|p{2cm}}
\hline
\hline
 Operation & Time taken \\
\hline\hline
Setup (1000 users) & 26.64 msec\\
\hline
Broadcast BT & 0.144 msec\\
\hline
FindRoute & 10.340 msec\\
\hline
FindRoute and Response  & 25.38 msec \\
\hline\hline
\end{tabular}
\end{table}
\par
Table ~\ref{tbl:bailout} shows the time for bailout operation after performing balance transfer. We recorded time for setting up the network from 1000 to 10000 users. Here the setup refers to establishing connections from landmark to all the other nodes in the network. For 1000 users, the setup time for Bailout is $0.0336$ seconds and the value increases with increase in number of users.. 'Create Edges' involves establishing edge and edge weights between the requester and other nodes who are willing to extend credits. The time taken for FindRoute is the total time taken for all the 10 nodes selected by landmark node to route to requester node. 
\begin{table}[htbp]
\centering
\caption {Bailout - Time recorded for operations} 
\label{tbl:bailout}
\begin{tabular}{p{2cm}|p{1.5cm}|p{1.5cm}|p{1.5cm}}
\hline
\hline
 Number of Users & Setup (time in msec) & FindRoute (time in msec) & Create Edges (time in msec) \\
\hline
1000 & 33.672& 0.072495& 3.63636\\
\hline
2000 & 58.514& 0.075099& 7.1581\\
\hline
4000  &123.363636 & 0.076168 & 16\\
\hline
8000  & 269.089109 &0.1378 & 42.297 \\
\hline
10000  & 369.6347 & 0.2779 & 52.7722 \\
\hline\hline
\end{tabular}
\end{table}
\par
\begin{figure} 
\centering
 \includegraphics[scale=0.42]{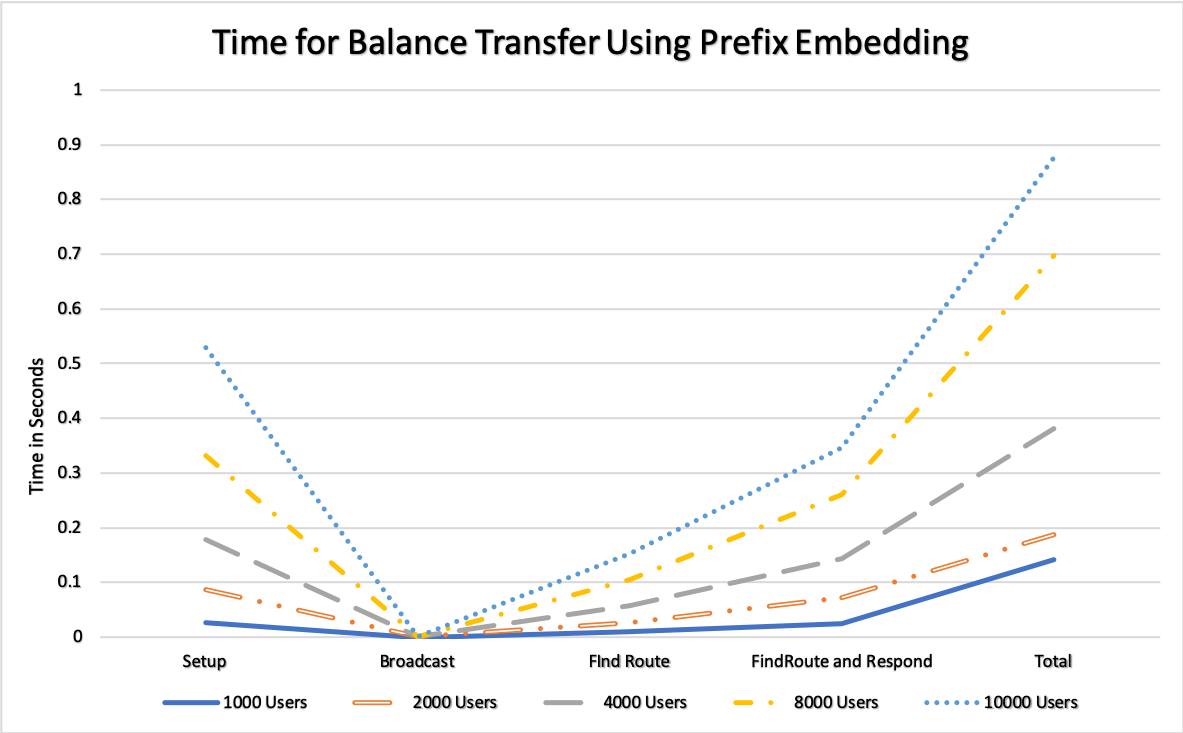}
\caption{Graph showing time taken for balance transfer using Prefix embedding. The time for all the operations increase with the number of users except Broadcast operation}
\label{fig:graph1}
\end{figure}

Figure~\ref{fig:graph1} shows the time taken for operations in balance transfer in prefix embedding. We obtain the values for 1000, 2000, 4000, 8000, 10000 users by taking a average on 100 values for each operations. We plotted the time taken for operation against number of users and the time for all the operations (except broadcast) with increasing number of users. The time taken for broadcast will not vary or increase with the number of users since, this operation involves creating a balance transfer tuple and sending it out into the network, regardless of the number of users in the network.  
 

%% file: conc.tex
\section{Conclusion and future work}
\label{sec:conc}

In this paper, we have proposed a new technique for rebalancing in credit networks that would help a poorly connected node rebalance its links, and become an active participant in the network, thus making the network robust, more competitive, and increasing the overall throughput of the network. Our method involves a poorly connected node creating incoming links using a process called \emph{balance transfer} and creating outgoing links using a process called \emph{bailout}. 
We present the high-level ideas and and prototype them in this paper; as a part of future work, we envision to implement the idea on a real world credit network such as Ripple~\cite{ripple}. 